\documentclass[aps,pra,groupaddress,showpacs,twocolumn]{revtex4-1}
\usepackage{amsfonts}
\usepackage{bm}
\usepackage{epsfig,amsmath,graphicx,amssymb,overpic}
\usepackage{color}

\def\be{\begin{equation}}
\def\ee{\end{equation}}
\def\bee{\begin{eqnarray}}
\def\ene{\end{eqnarray}}
\def\bes{\begin{subequations}}
\def\ees{\end{subequations}}

\newcommand {\tmu}{\tilde{\mu}}
\newcommand {\tphi}{\tilde{\phi}}

\newcommand{\p}{{\cal P}}
\newcommand{\PT}{{\cal PT}}
\newcommand{\T}{{\cal T}}
\newcommand{\eps}{ \epsilon}

\begin{document}

\title{Solitons in a nonlinear Schr\"odinger equation with $\PT$-symmetric potentials and inhomogeneous nonlinearity: stability and excitation of nonlinear modes}
\author{Zhenya Yan$^{1}$}
\email{zyyan@mmrc.iss.ac.cn}
\author{Zichao Wen$^{1}$}
\author{Vladimir V. Konotop$^{2}$}
\affiliation{$^{1}$Key Laboratory of Mathematics Mechanization, Institute of Systems
Science, AMSS, Chinese Academy of Sciences, Beijing 100190, China \\
$^{2}$Centro de F\'{i}sica Te\'orica e Computacional and Departamento
de F\'{i}sica,  Faculdade de Ci\^encias, Universidade de Lisboa,  Campo Grande, Edif\'icio C8, Lisboa 1749-016,  Portugal
}

\begin{abstract}
We report branches of explicit expressions for nonlinear modes in parity-time ($\PT$) symmetric potentials of several types. For the  single-well and double-well potentials the found solutions are two-parametric and appear to be  stable even when the $\PT$-symmetry of respective underlying linear models is broken. Based on the examples of these solutions we describe an algorithm of excitation of a stable nonlinear mode in a model, whose linear limit is unstable. The method is based on the adiabatic change of the control parameter driving the mode along a branch bifurcating from a stable linear mode. The suggested algorithm is confirmed by extensive numerical simulations.
\end{abstract}

\pacs{42.65.Tg, 42.65.Jx, 42.65.Wi, 05.45.Yv,  11.30.Er \\  PHYSICAL REVIEW A 92, 023821 (2015)}

\maketitle

\section{Introduction.}

A common practical requirement for a nonlinear system to have localized solutions is the stability of the zero background. This requirement on the one hand ensures absence of growing small fluctuations far from the nonlinear mode. More importantly, this means the stable existence of a system itself without any excitations, i.e., in the ``vacuum state", in which the system is prepared experimentally and against which nonlinear modes are excited. In the conservative case this implies possibility for stable propagation of linear waves (real eigenvalues of the linear Hamiltonian), while in dissipative systems this means decay of all small amplitude excitations (the background is an attractor with a nonzero basin). In this context parity ($\p$) -- time ($\T$) symmetric~\cite{Bender98,Bender2} systems represent a special case which on the one hand obey gains and losses, and  one the other hand may have pure real spectra in some domains of the parameter spaces~\cite{Bender2} (this situation is referred to as unbroken $\PT$-symmetric phase~\cite{Bender98}) allowing for propagation of linear waves and possessing continuous families of solutions. Therefore linear potentials like parabolic~\cite{Kato},
Scarff II~\cite{sh}, or $\PT$-symmetric extension of Rosen-Morse II potential~\cite{Levai} obeying pure real spectra and supporting localized modes,  received particular attention also from the point of view of the existence of nonlinear families; see Ref.~\cite{ZK}, Refs.~\cite{Musslimani,SII}, and Ref.~\cite{RMII}, respectively.

Nonlinear modes can also be found in a region where the linear $\PT$-symmetry is broken~\cite{Musslimani}. Moreover, families of nonlinear modes can be stable~\cite{ZK_PRL,Segev} in those regions. However, the mathematical existence of such  nonlinear modes, and of even stable ones,  does not yet guarantee their practical utility, because the way of their excitation in a system which is linearly unstable in its ``vacuum" state, remains questionable. This leads us to the first goal of the present paper, which is a suggestion on how {\em stable nonlinear modes can be excited in systems where the linear $\PT$-symmetry is broken}. The idea is based on the possibility of ``switching-on" nonlinearity simultaneously with gain and dissipation. Such possibility can be implemented, in particular, when the nonlinearity and gain-and-loss strength are characterized by a single parameter (let us call it $\epsilon$) and disappear when this parameters becomes zero ($\epsilon=0$). If at $\epsilon=0$ the system is Hamiltonian, it allows for stable propagation of the linear modes, and the only stability issue which has to be verified is the stability of the solution branch $\epsilon>0$, bifurcating from $\epsilon=0$. Then, if the stability is confirmed, one can consider the adiabatic grows of $\epsilon=\epsilon(t)$ in time as a way to excite a nonlinear mode, which persists stable even if the final value of $\epsilon$ corresponds to the underline linear system with broken $\PT$-symmetric phase.

To solve this problem mathematically, a suitable framework is the use of potentials supporting some exact solutions. Such complex potentials can be constructed, say using the ``inverse engineering" as this was suggested in~\cite{AKSY} (here we also mention other examples of particular exact solutions for $\PT$-symmetric potentials published in~\cite{Musslimani,HP}). The method consists in computing a complex potential starting with an {\em a priori} given solution, which however must satisfy some constraints to ensure the existence of the potential. Generally speaking, the potentials supporting particular exact solutions appear of rather sophisticated forms, what may constitute a significant difficulty for their practical implementation. Therefore as a complementary task of this paper we consider generation of  the modes in potentials of relatively simple and experimentally feasible forms, bearing in mind their applications in optics of atomic gases ~\cite{gases} as well as in the $\PT$-symmetric physics of Bose-Einstein condensates~\cite{BECs,Four-well,Cart}, where $\PT$-symmetric potentials can be created and modified {\it in situ}.

The rest of the paper is organized as follows. In Sec.~\ref{sec:Gen}, we describe the general theory and approach to the nonlinear Schr\"odinger  equation with the $\mathcal{PT}$-symmetric potentials allowing for exact particular solutions. In Sec.~\ref{sec:parabol} we discuss applications of the method to the parabolic potential and in Secs.~\ref{subs1} and \ref{subs2} we consider linear and nonlinear modes in both single-well and double-well potentials, respectively. Particularly, the problem of the nonlinear mode excitations in the single-well and double-well $\PT$-symmetric potentials allowing for particular exact solutions is also addressed in Secs. ~\ref{subs1} and ~\ref{subs2}. Our results are summarized in the Conclusion.

 \section{General theory and approach}
 \label{sec:Gen}

We consider the nonlinear Schr\"odinger (NLS) equation with the $\mathcal{PT}$-symmetric potential and space-modulated nonlinearity (abbreviated below as $\mathcal{PT}$-NLSE)
\bee\label{nls}
 i\partial_t \psi=-\frac12\partial^2_x\psi+U_\eps (x)\psi+G_\eps (x)|\psi|^2\psi,
\ene
where $\partial_t=\partial/\partial t,\, \partial_x=\partial/\partial x$, $\psi=\psi(x,t)$ is the complex envelope of
the electrical field, $U_\eps(x)=U_\eps^*(-x)$ (i.e., ${\rm Re}[U_\eps(x)]={\rm Re}[U_\eps(-x)]$,\, ${\rm Im}[U_\eps(x)]=-{\rm Im}[U_\eps(-x)]$) and $G_\eps(x)$ describe the complex-valued linear $\PT$-symmetric potential and real-valued inhomogeneous nonlinearity, respectively, and an asterisk stands for the complex conjugation. For specific the real-valued potential, i.e., if $U_\eps(x)=U_\eps^*(x)$, Eq.~(\ref{nls}) reduces to the conservative NLS equation with space-modulated linear and nonlinearities coefficients. In that case, particular exact solutions and  corresponding dynamical behaviors were extensively studied in literatures (see, e.g.,~\cite{hm, bb08, yan10, yan09, yan12}).

In order to implement the procedure described in the Introduction we assume that:
\begin{eqnarray}
\label{U}
U_\eps(x)=\sum_{j=0}^2\epsilon^{2j}V_j(x)+i\epsilon W(x),
\\
\label{G}
G_\eps (x)=\epsilon^2 G(x),
\end{eqnarray}
where $V_j(x)\, (j=0,1,2)$ are constituents of the real potential, $W(x)$ is the real gain-and-loss distribution, $\eps\geq 0$ is the bifurcation parameter parameterizing a branch of the solutions: we emphasize that generally speaking $\epsilon $ is not considered small. To ensure the $\PT$-symmetry we consider $V_j(x)=V_j(-x)\, (j=0,1,2)$ and $W(x)=-W(-x)$. At $\eps=0$, Eq.~(\ref{nls}) becomes the linear Schr\"odinger equation. Our main interest will be focused on the case where $W(x) \not\equiv 0$ (i.e. when the linear potential $U_\eps(x)$ is non-Hermitian).

We concentrate on stationary solutions of $\mathcal{PT}$-NLSE (\ref{nls}) in the form $\psi(x,t)=\phi(x)e^{-i\mu t} $, where $\mu$ is a real spectral parameter, and the complex valued nonlinear eigenmode $\phi(x)$ satisfies the stationary NLS equation
\bee\label{snls}
\mu \phi=-\frac12\frac{d^2\phi}{dx^2}+U_\eps (x)\phi+G_\eps  (x)|\phi|^2\phi,
\ene
subject to the zero boundary conditions $\lim_{x\to\pm\infty}\phi(x)=0$.

Now, following the strategy described in the Introduction we assume that the solution of the linear eigenvalue problem
\begin{eqnarray}
\label{lin}
L_0\tphi_n(x)= \tmu_n \tphi_n(x), \quad  L_0=-\frac 12 \frac{d^2}{dx^2}+V_0(x),
\end{eqnarray}
where $n=0,1,2,...,$ is known. In Eq.~(\ref{lin}) we assume also that the spectrum is discrete (what is applied for all examples considered in this paper) and distinguished the  eigenfunctions and eigenvalues of this linear problem  by tildes. In order to obtain a nonlinear branch of solutions $\phi_n$ which bifurcates from  $\tphi_n(x)$  we require $\lim_{x\to\pm\infty}\tphi_n(x)=0$. Thus all $\tphi_n(x)$ can be considered real without loss of generality. Since $L_0$ is Hermitian, all the eigenvalues $\tmu_n$ are also real, what means the stability of the respective linear system (here we exclude the situation where zero is an eigenvalue of $L_0$).

Turning now to $\eps>0$ we observe that in the presence of gain-and-loss distribution, stationary modes (if any) must have non-zero (hydrodynamic) current, i.e., $x$-dependent argument. Respectively we consider the construction of the $\PT$-symmetric potential and nonlinearity for the modes having the form
\begin{eqnarray}
\label{ansatz_0}
\phi_n(x)= 
\tphi_n(x,\epsilon)\exp\left(i\epsilon\!\!\int_{-\infty}^{x} v_n(\xi) d\xi\right),  \quad
\end{eqnarray}
where the real function $v_n(x)$ is the hydrodynamic velocity and $n$ stands here for the identification of the family bifurcating from the $n$-th linear eigenstate.  Since this ansatz implies that the modulus of the nonlinear modes persists equal to the linear distribution, a possibility of constructing such modes is not obvious. In order to address this issue we substitute Eq.~(\ref{ansatz_0}) in Eq.~(\ref{snls}) and obtain relations linking the phase:
\begin{eqnarray}
\label{W}
\frac{d}{dx}\left(v_n(x)|\tphi_n(x,\epsilon)|^2\right)=2W(x)|\tphi_n(x,\epsilon)|^2,
\end{eqnarray}
and the amplitude:
\begin{eqnarray}
 \label{V_gen}
\frac12 v_n^2(x)+ G(x)|\tphi_n(x,\epsilon)|^2+V_1(x)+\epsilon^2V_2(x)=\nu_n,
\end{eqnarray}
where we introduce the shift of the eigenvalue $\nu_n$ determined by $\mu-\tmu_n=\eps^2 \nu_n$.

The last equation has a free parameter $\epsilon$ which leaves much freedom in constructing a solution. Although we do not require $\epsilon$ to be small,  we nevertheless restrict further consideration to two basic cases as follows:

{\em Case 1.} The amplitude of $\phi_n(x)$ is independent on $\epsilon$, i.e. $\tphi_n(x,\epsilon)=\tphi_n(x)$ and
\begin{eqnarray}
\label{V1}
\frac12 v_n^2(x)+G(x)\tphi_n^2(x)+V_1(x)=\nu_n, \,\, V_2(x)\equiv 0.
\end{eqnarray}

{\em Case 2.} The amplitude of $\phi_n(x)$ is proportional to $\epsilon$, i.e. $\tphi_n(x,\epsilon)=\epsilon \tphi_n(x)$ and
\begin{eqnarray}
\label{V2}
\frac12 v_n^2(x)+V_1(x)=\nu_n, \,\, G(x)\tphi_n^2(x)+V_2(x)=0.
\end{eqnarray}

The obtained system (\ref{W}) with Eq.~(\ref{V_gen}) [cf. Eq.~(\ref{V1}) or Eq.~(\ref{V2})] still contains freedom in the definition of the potentials and  the hydrodynamic velocity. Therefore we impose further constraints on the part linear $V_j(x)$ for $j\geq 1$, $W(x)$ and nonlinear $G(x)$ potentials, as well as the hydrodynamic velocity $v_n(x)$, requiring them to be localized, i.e.,  $V_j(x)\to 0$ for $j\geq 1$, $W(x)\to 0$, $G(x)\to 0$, and $v_n(x)\to 0$ at $|x|\to \infty$.  Considering now Eq.~(\ref{V_gen}) in the limit $|x|\to \infty$ we readily obtain that $\nu_n=0$. Thus, in this case a nonlinear mode has not only the same form but also the same eigenvalue as its linear counterpart at $\epsilon=0$ does.

Thus we have obtained the profiles of the nonlinear modes in an exact analytical form $\psi_n(x,t)=\phi_n(x)e^{-i\mu_n t}$ with $\phi_n(x)$ being given by Eq.~(\ref{ansatz_0}). Our next main task is to study the stability of the nonlinear modes. We do this numerically by two standard approaches. First, we address the linear stability of a nonlinear mode $\psi_n(x,t)=\phi_n(x)e^{-i\mu_n t}$, employing the ansatz
\begin{equation}
\label{perturbation}
\psi_n(x,t)=\left\{\phi_n(x)+\varrho \left[f(x)e^{-i\delta t}+g^*(x)e^{i\delta^* t}\right]\right\}e^{-i\mu_n t},\,\,\,\,
\end{equation}
where $\varrho\ll 1$, and $f(x)$ and $g(x)$ are the eigenfunctions of the linear eigenvalue problem:
\begin{eqnarray}
\left(\begin{array}{cc}   L_\eps & G_\eps (x)\phi_n^2(x) \vspace{0.1in}\\   -G_\eps(x)\phi_n^{*2}(x) & -L_\eps^* \\  \end{array}\right)
\left(  \begin{array}{c}    f \\    g \\  \end{array} \right)
=\delta \left(  \begin{array}{c}   f\\    g \\  \end{array}\right) \qquad
\end{eqnarray}
with
\begin{eqnarray}
L_\eps=-\frac{1}{2}\partial^2_x+U_{\eps}(x)+2 G_\eps(x)|\phi_n(x)|^2-\mu_n.
\end{eqnarray}
 The solution is linearly unstable if $\delta$ has a non-zero imaginary part, otherwise it is stable.

Second, we test the stability by direct propagation using an exact solution $\psi_n(x,t)=\phi_n(x)e^{-i\mu_n t}$ with $\phi_n(x)$ given by Eq.~(\ref{ansatz_0})  with a noise perturbation of order $1\%$ of the initial amplitude $|\psi_n(x,0)|$,  as the initial condition for Eq.~(\ref{nls}).

\section{Perturbed $\PT$-symmetric linear parabolic potential}
\label{sec:parabol}

Below we concentrate on the physically relevant case of the parabolic (harmonic) potential $V_0(x)=\omega^2x^2/2$.  Without loss of generality, one can scale out the frequency $\omega$ making it one to yield:
\begin{eqnarray}
\label{hp}
V_0(x)=\frac12 x^2.
\end{eqnarray}
The families of the nonlinear modes of this potential at $W(x)=x$, $V_j(x)\equiv 0\, (j=1,2)$ and constant nonlinearity $G(x)\equiv$const were considered in~\cite{ZK}.

Now the profile of linear modes related to Eq.~(\ref{lin}) is described by the Gauss-Hermite functions
\begin{eqnarray}
\label{Hermite}
\tphi_n(x)=H_n(x)e^{-x^2/2},
\end{eqnarray}
where $H_{n}(x)=(-1)^{n}e^{x^{2}}(d^{n}e^{-x^{2}})/(dx^{n})$  is the Hermite polynomial with $n=0,1,2,...$, and the eigenvalue is
\begin{eqnarray}
\label{mun}
\mu=\tmu_n= n+\frac12,
\end{eqnarray}
in which we have $\nu_n=0$.

We will also concentrate on the specific Gaussian form of the nonlinearity
\bee
\label{poten3}
G(x)=2\sigma e^{-\alpha x^2}
\ene
with the characteristic width $1/\sqrt{\alpha}$ with $\alpha>0$ and constant $\sigma$ (particularly, the nonlinearity is a constant for $\alpha=0$). This is a natural choice, for example, for the Bose-Einstein applications, where nonlinearity can be controlled through the optical Feshbach resonance, with $G(x)$ describing the profile of the laser beam (see, e.g., \cite{bec}).

Thus, the solution of the problem is reduced to two steps as follows. First, given the gain-and-loss distribution $W(x)$ from Eq.~(\ref{W}) one obtains the hydrodynamic velocity:
\begin{eqnarray}
\label{vn}
v_n(x)=\frac{2}{|\tphi_n(x,\eps)|^2}\int_{-\infty}^x W(\xi)|\tphi_n(\xi,\eps)|^2 d\xi.
\end{eqnarray}
Second, the ``correction" to the conservative part of the Hamiltonian is computed from Eqs.~(\ref{V1})-(\ref{V2}),
%
with  $G(x),\, v_n(x)$, and $\tphi_n(x)$ given by Eqs.~(\ref{poten3}), (\ref{vn}), and (\ref{Hermite}).

In the meantime, formula (\ref{vn}) imposes the additional constraint on the choice of the imaginary potential $W(x)$, which must ensure the existence of the integral in Eq.~(\ref{vn}). This problem can be overcome if one again uses the inverse engineering, i.e. considers the hydrodynamic velocity given and finds $W(x)$ from Eq.~(\ref{W}).

\section{A single-well potential}
\label{subs1}

\subsection{$\PT$-symmetry phases of the linear problem}

The simplest single-well potential is obtained by setting $V_j(x)\equiv 0\,\, (j=1,2)$, which corresponds to the Case 1 in Eq.~(\ref{V1}) [in the Case 2 in Eq.~(\ref{V2}) one has $G(x)\tphi_n^2(x)\equiv 0$, which is the trivial solution]. Recalling that $\tmu_n$ is given by Eq.~(\ref{mun}) (i.e., $\nu_n=0$) and considering $\tphi_n(x)$ given by Eq.~(\ref{Hermite}) we readily conclude that such a choice is possible only for attractive (focusing) nonlinearities $\sigma<0$ (without loss of generality we set $\sigma=-1$) for which the hydrodynamic velocity reads
\bee\label{vn2}
 v_n(x)=2H_n(x)e^{-(\alpha+1) x^2/2}.
\ene
Now from Eq.~(\ref{W}) we can find the gain-and-loss distributions (i.e. imaginary part of the potential)
\begin{equation}
 \label{poten2}
 W_n(x)=[6nH_{n\!-\! 1}(x)\!-\!(\alpha\!+\!3)xH_n(x)]e^{-(\alpha\!+\!1)x^2/2}.
\end{equation}
In order $W_n(x)$ to be an odd function, i.e. to support the $\PT$-symmetry,  we require $n$ to be an even number: $n=0,2,4,...,$. The functions $W_n(x)$ with odd $n$ are even and do not satisfy the condition of the $\PT$-symmetry. We do not consider them here but observe that for the odd $n$ Eq.~(\ref{snls}) still obeys exactly localized solutions of the form  (\ref{ansatz_0}) with $V_0(x)$,\, $\tphi_n(x)$,\, $G(x)$,\, $v_n(x)$, and $W_n(x)$ given by Eqs.~(\ref{hp}), (\ref{Hermite}), (\ref{poten3}),  (\ref{vn2}), and (\ref{poten2}), respectively.

We also observe that all the members of the $W_n(x)$ family of potentials (i.e. the potentials corresponding to different $n$) are {\em two}-parametric, i.e. are determined by the amplitude [it is given by  $\epsilon$ when substituted in $U_\eps(x)$] and by the internal parameter $\alpha$.

In order to establish domains of unbroken $\PT$-symmetry phase of the linear $\PT$-symmetric potential $U_\eps$ for different $n$  we address the spectral problem
\begin{eqnarray}
\label{linearPT}
 \hat L_n\Psi(x)=\lambda_n\Psi(x), \quad \hat L_n= L_0 + i\eps W_n(x),
\end{eqnarray}
where $L_0$ is given by Eq.~(\ref{lin}), $\lambda_n$ and $\Psi(x)$ are eigenvalues and eigenfunctions, respectively. Since the discrete spectrum of a $\PT$-symmetric potential is either real or appears in complex conjugated pairs, we conclude there exists a nonzero domain of the parameter $\epsilon$ for which the $\PT$-symmetry remains unbroken.

\begin{figure}
 	\begin{center}
 	 	\hspace{-0.1in}{\scalebox{0.3}[0.3]{\includegraphics{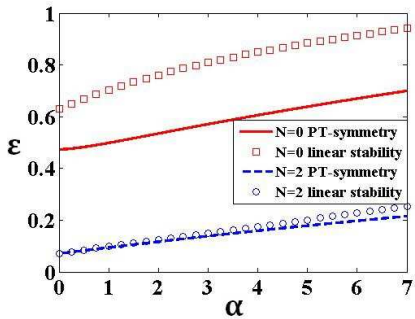}}}
 	\end{center}
 	\vspace{-0.15in} \caption{\small (Color online). Solid (red) and dashed (blue) curves indicate the lines of phase transitions for the $\PT$-symmetric potentials (\ref{W0}) and (\ref{W1}) with $n=0,\,2$. The unbroken (broken) $\PT$-symmetric phase is in the domain below (above) the phase breaking lines. Square-shaped (red) and circle-shaped (blue) curves indicate the lines of linear stability of solitons (\ref{eq:nonlin_mod_1well}) with $n=0,\,2$. The stable (unstable) soliton is in the domain below (above) the linear stable lines. }
 	\label{fig1n}
 \end{figure}

The simplest potentials with nonzero complex parts are given by $n=0$:
\begin{eqnarray}
\label{W0}
U_\eps(x)=\frac{x^2}{2}-i\eps (\alpha+3) xe^{-(\alpha+1)x^2/2} \qquad\qquad\qquad \qquad
\end{eqnarray}
and $n=2$:
\begin{equation}
\label{W1}
U_\eps(x)\!=\!\frac{x^2}{2}-2i\eps x [2(\alpha+3)x^2\!-\!(\alpha+15)]e^{-(\alpha+1)x^2/2}.
\end{equation}
In Fig.~\ref{fig1n} we show the domains of broken and unbroken phase on the  $(\alpha,\eps)$-plane. The both curves in the figure grow with $\alpha$, what can be understood from the fact that growth of $\alpha$ corresponds to the shrinking of the gain-and-loss domains.

\begin{figure}
	\begin{center}
		\hspace{-0.05in}{\scalebox{0.43}[0.5]{\includegraphics{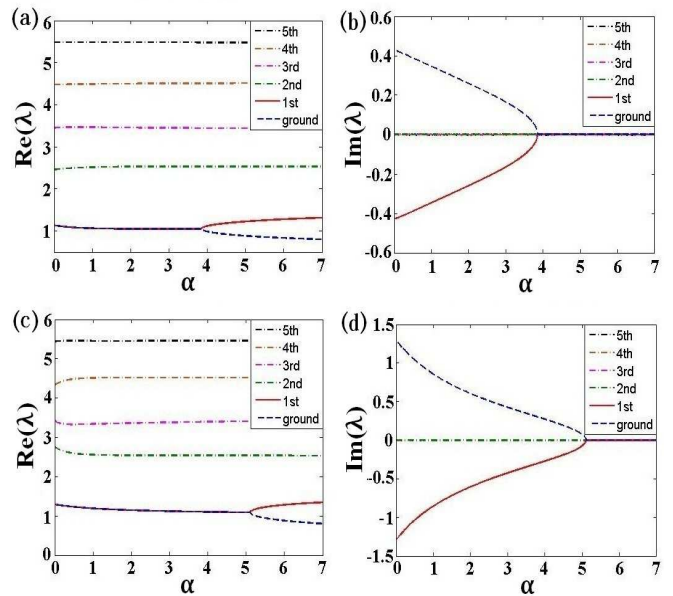}}}
				\end{center}
	\vspace{-0.15in}
	\caption{ (color online).  Real (a), (c) and imaginary (b), (d) parts of the eigenvalues $\lambda_n$ [see Eq.~(\ref{linearPT})] as functions of $\alpha$ for the potential (\ref{W0}) at $n=0,\, \epsilon=0.6$ (the upper row) and for the potential (\ref{W1}) at $n=2,\, \epsilon=0.18$ (the lower row).}		 \label{fig1i}
\end{figure}

In both cases, illustrated in Fig.~\ref{fig1i}, the spontaneous symmetry breaking occurs due to collision of the two lowest states  as $\alpha$ decreases (what corresponds to the increase of the width of the gain-and-loss domains). All upper eigenstates (we checked numerically for the 6 lowest states) remain real. This can be understood from the fact that for the levels with large $n$ the imaginary part of the potential represents a weak perturbation while the lowest levels are the most strongly deformed ones. We also notice that the instability is oscillatory (i.e. the two emergent complex eigenvalues have nonzero real part).

\subsection{Nonlinear modes in a single-well potential}

\begin{figure}
	\begin{center}
		\hspace{-0.05in}{\scalebox{0.42}[0.45]{\includegraphics{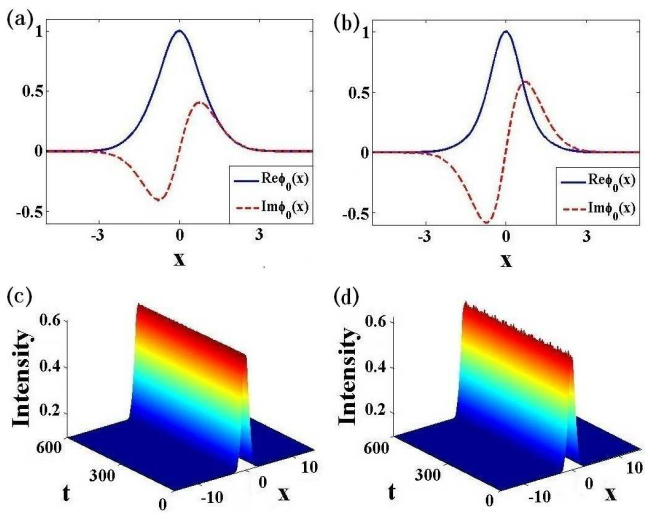}}}
	\end{center}
	\vspace{-0.15in} \caption{(Color online).
 One-hump nonlinear modes given by Eq.~(\ref{eq:nonlin_mod_1well}) with $n=0$ and $\alpha=1$ for (a) $\epsilon=0.45$
 (real spectrum of the operator $\hat{L}_0$ , i.e. unbroken linear $\PT$-symmetry) and (b) $\epsilon=0.7$ (broken linear $\PT$-symmetry). (c) Stable and (d) periodically varying propagation of the nonlinear modes described by Eq.~(\ref{eq:nonlin_mod_1well}) and corresponding to the weakly perturbed initial conditions shown in (a) and (b), respectively.
 } \label{fig3}
\end{figure}

Now we turn to the nonlinear modes in the $\PT$-symmetric potentials (\ref{W0}) and (\ref{W1}). The expression for the nonlinear modes is obtained form Eqs.~(\ref{ansatz_0}) and (\ref{vn2}):
\begin{equation}
\label{eq:nonlin_mod_1well}
\phi_n(x)=H_n(x)e^{-x^2/2}\!\exp\!\left(\!2i\epsilon\!\!\int_{-\infty}^{x}\!\!\!H_n(\xi)e^{-(\alpha+1)\xi^2/2}d\xi \right).
\end{equation}
Two examples  of the modes are illustrated in Fig.~\ref{fig3} and their linear stability analysis is presented Fig.~\ref{fig1n}. The feature, most relevant to the present consideration, is displayed by the domains between the solid line and squares for $n=0$ and between the dashed line and circles for $n=2$. In these domains the nonlinear modes [given by Eq.~\eqref{eq:nonlin_mod_1well} with $n=0$ and $n=2$] are stable, while the respective  linear $\PT$--symmetric phases are broken, i.e. linear stability of the nonlinear modes is extended beyond the unbroken linear $\PT$-symmetric phase. Below we explore these domains in order to ``draw" the mode along the branch bifurcating from the linearly stable mode. Before that, however we show the check of stability by means of the direct propagation of the initially stationary state Eq.~(\ref{eq:nonlin_mod_1well}) with a noise perturbation of order $1\%$. In Fig.~\ref{fig3} (c) we show stable propagation of the soliton for the parameters belonging to the domain of the unbroken linear $\PT$-symmetric phase  of  the operator $\hat L_0$ [defined in Eq.~(\ref{linearPT})], and to the linearly stable nonlinear mode. In Fig.~\ref{fig3} (d) we illustrate the evolution of the mode where  the linear $\PT$-symmetric phase is broken and the nonlinear mode is linearly stable. In this last case we observe the oscillatory (breather-like behavior).

\begin{figure}
	\begin{center}
		\hspace{-0.05in}{\scalebox{0.42}[0.45]{\includegraphics{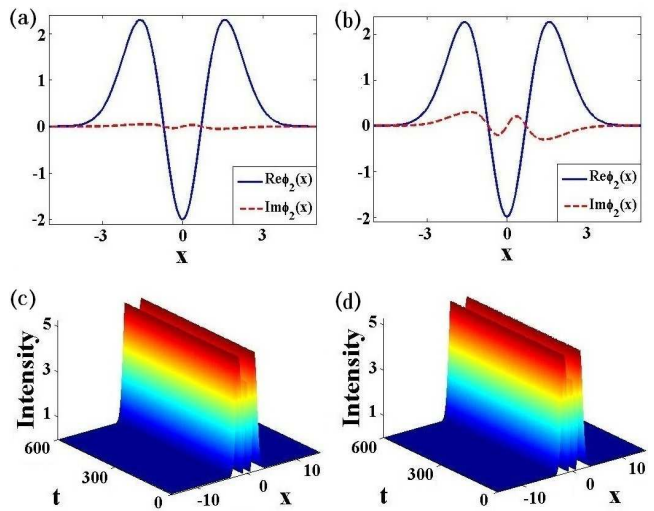}}}
	\end{center}
	\vspace{-0.15in} \caption{(Color online).
		 Two-hump nonlinear modes given by Eq.~(\ref{eq:nonlin_mod_1well}) with $n=2$ and $\alpha=2$ for (a) $\epsilon=0.02$ (real spectrum of the operator $\hat{L}_2$ , i.e. unbroken linear $\PT$-symmetry) and (b) $\epsilon=0.12$ (broken linear $\PT$-symmetry). Both nonlinear modes are linearly stable [see Fig.~\ref{fig1n}] what are illustrated in panels (c) and (d) where the direct numerical simulations of Eq.~(\ref{eq:nonlin_mod_1well})  are performed with weakly perturbed initial profiles (a) and (b).} \label{fig4}
\end{figure}

Similarly, Figs.~\ref{fig4} (a) and (c) display the initial states and stable intensity evolution of a two-hump solitary wave [it is described
by Eq.~\eqref{eq:nonlin_mod_1well} with $n=2$] for the parameters which guarantee both the real spectrum of the operator $\hat L_2$ (unbroken $\PT$-symmetric phase) as well as the linear stability of the nonlinear mode. Meantime, Figs.~\ref{fig4} (b) and (d) show a two-hump soliton for the parameters corresponding to broken liner $\PT$-symmetry but still stable the nonlinear mode. In both numerical simulations we observed robustness of the nonlinear modes with respect to weak initial noise.

\begin{figure}[!tp]
	\begin{center}
		\vspace{0.1in}{\scalebox{0.42}[0.5]{\includegraphics{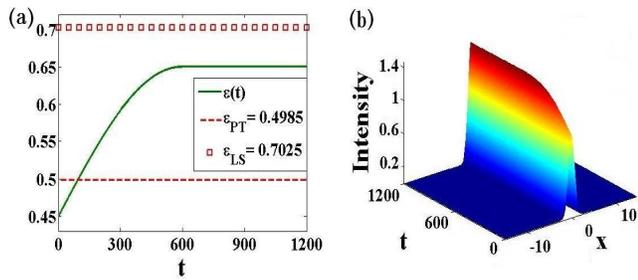}}}
		\\
		
	\end{center}
	\vspace{-0.15in} \caption{ (Color online). (a) Time dependence of the parameter $\epsilon(t)$ (solid line) given by Eq.~(\ref{swell0}). The squares and solid line (they correspond to the squares and dashed line in Fig.~\ref{fig1n}) indicate the phase $\PT$-symmetry breaking phase transition in the linear case and loss of the stability of the nonlinear modes, respectively. (b) Intensity evolution of a nonlinear mode governed by Eq.~(\ref{nls-d}) with the initial condition $\psi_0(x,t=0)=\phi_0(x)$ given by Eq.~(\ref{eq:nonlin_mod_1well}) at $\epsilon=0.45$ for the single-well potential $V_0(x)$ (\ref{hp}) and gain-and-loss distribution $W_0(x)$ (\ref{poten2}). Other parameters are $n=0$ and $\alpha=1$.}
	\label{fig5-swell-t}
\end{figure}

Now we turn to the excitation of nonlinear modes by means of slow change of the control parameter $\epsilon(t)$ which is now considered as a function of time. More specifically we consider simultaneous adiabatic switching-on the gain-and-loss distribution and the nonlinearity, modeled by [cf. Eq.~(\ref{nls})]
 \begin{equation}
 i\partial_t\psi=-\frac12\partial_x^2\psi+[V_0 (x)+i\epsilon(t)W_n(x)]\psi
 +\epsilon^2(t)G(x)|\psi|^2\psi,
\label{nls-d}
\end{equation}
where the single-well potential $V(x)$, nonlinearity $G(x)$ and gain-and-loss distribution $W(x)$ are given by Eqs.~(\ref{hp}), (\ref{poten3}) and (\ref{poten2}), and
\begin{eqnarray}
\label{swell0}
\epsilon(t)=\left\{
\begin{array}{ll} 0.2\sin\left(\dfrac{\pi t}{1200}\right)+0.45, & 0\leq t<600
\vspace{0.1in}\\
0.65, & 600\leq t\leq 1200. \quad
\end{array}
\right.
\end{eqnarray}
This choice of the final value of $\epsilon$ is justified by the fact that the whole ``trajectory" $\epsilon(t)$, it is shown in Fig.~\ref{fig5-swell-t}(a), belongs to the parameter domain where the nonlinear mode is stable.

Fig.~\ref{fig5-swell-t}(b) exhibits the evolution of the solution $\psi(x,t)$ governed by Eqs.~(\ref{nls-d}) and (\ref{swell0}) subject to the initial condition given by (\ref{eq:nonlin_mod_1well}) with $n=0$, i.e., for the single-well potential $V_0(x)$, nonlinearity $G(x)$ and gain-and-loss distribution $W_0(x)$ given by Eqs.~(\ref{hp}), (\ref{poten3}), and (\ref{poten2}), respectively. One observes remarkably stable propagation with the increasing amplitude of nonlinear modes, which is driven from a nonlinear mode at the system parameters of unbroken linear $\PT$-symmetric phase, to the stable nonlinear mode at the parameters where the liner $\PT$-symmetric phase is broken (i.e., while $\epsilon(t)$ is growing there occurs the linear $\PT$-symmetry phase transition and linear modes becomes unstable).

\begin{figure}[!tp]
 	\begin{center}
 		\vspace{0.1in}
 \hspace{0.15in}{\scalebox{0.42}[0.5]{\includegraphics{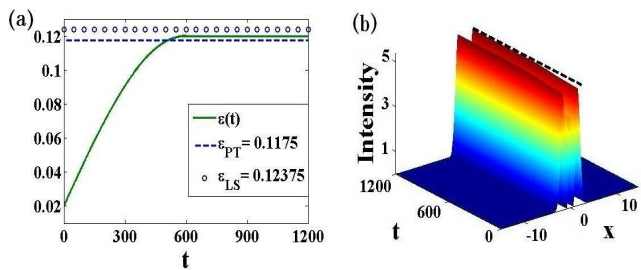}}}
 	\end{center}
 	\vspace{-0.15in} \caption{(Color online) (a) Time dependence of the parameter $\epsilon(t)$ (solid line) given by Eq.~(\ref{swell2}). The circles and dashed line are explained in Fig.~\ref{fig1n}.
 		(b) Intensity evolution of a two-hump nonlinear mode of Eq.~(\ref{nls-d}) with $n=2$, and $\alpha=2$ subject to the initial condition $\psi_2(x,t=0)=\phi_2(x)$ given by Eq.~(\ref{eq:nonlin_mod_1well}) at $\epsilon=0.02$ for the single-well potential $V_0(x)$ (\ref{hp}) and gain-and-loss distribution $W_2(x)$ (\ref{poten2}). The dashed line shows the finally established amplitude of the mode which is different.
 		}
 	\label{fig6-swell-n2-t}
 \end{figure}

Next we consider the excitation of the mode (\ref{eq:nonlin_mod_1well}) with $n=2$ described by Eq.~(\ref{nls-d}) with the control parameter adiabatically changing according to the law [see Fig.~\ref{fig6-swell-n2-t} (a)]
\begin{eqnarray}\label{swell2}
\epsilon(t)=\left\{
\begin{array}{ll} 0.1\sin\left(\dfrac{\pi t}{1200}\right)+0.02, & 0\leq t<600, \vspace{0.1in}\\
0.12, & 600\leq t\leq 1200.
\end{array}
\right. \quad
\end{eqnarray}
The single-well potential $V_0(x)$, nonlinearity $G(x)$ and gain-and-loss distribution $W_2(x)$ given by Eqs.~(\ref{hp}), (\ref{poten3}) and (\ref{poten2}), respectively.

Like in the previous case we observe stable evolution of the nonlinear mode ``prepared" in a system with unbroken $\PT$-symmetric phase and drawn to the system with broken $\PT$-symmetric phase. The amplitude of the nonlinear mode grows with $\epsilon(t)$ and remains unchanged after the control parameter reaches its final value [see Fig.~\ref{fig6-swell-n2-t} (b)].

\section{A double-well potential}	
\label{subs2}
\subsection{$\PT$-symmetry phases of the linear problem}

Now we consider multi-well
potential with $V_1(x)V_2(x)\not\equiv 0$, which corresponds to the Case 2 in Eq.~(\ref{V2}). Recalling that $\tmu_n$ is given by Eq.~(\ref{mun})  and considering $\tphi_n(x)$ given by Eq.~(\ref{Hermite}) we  conclude that now the  hydrodynamic velocity takes the form
\bee\label{vn2d}
 v_n(x)=H_n(x)e^{-x^2/2}
\ene
(i.e., it is now fixed having no free parameters).

From Eqs.~(\ref{W}) and (\ref{V2}) it follows that the gain-and-loss distributions (i.e. imaginary potentials which are odd functions) numbered by an even number $n$, i.e. ensuring $\PT$-symmetry of the potential, and generated by the velocity field (\ref{vn2d}) are given by
\begin{equation}
 \label{poten2d}
  W_n(x)=\left[3nH_{n-1}(x)-\frac{3}{2}xH_n(x)\right]e^{-x^2/2},
\end{equation}
The real parts of the respective linear potential read
\bee \label{v1n}
 V_{1n}(x)=-\frac12H_n^2(x)e^{-x^2}, \,\,\, \qquad \\
   V_{2n}(x)=-2\sigma H_n^2(x)e^{-(\alpha+1)x^2}.
\label{v2n}
\ene
All the members of the $W_n(x)$ family of potentials are two-parametric; they are determined by the amplitude given by  $\epsilon $ in $U_\eps(x)$, and by the internal parameter $\alpha$ introduced in  Eq.~(\ref{v2n}).


\begin{figure*}[!tp]
 	\begin{center}
 	\vspace{0.1in}
 	\hspace{-0.15in}{\scalebox{0.8}[0.8]{\includegraphics{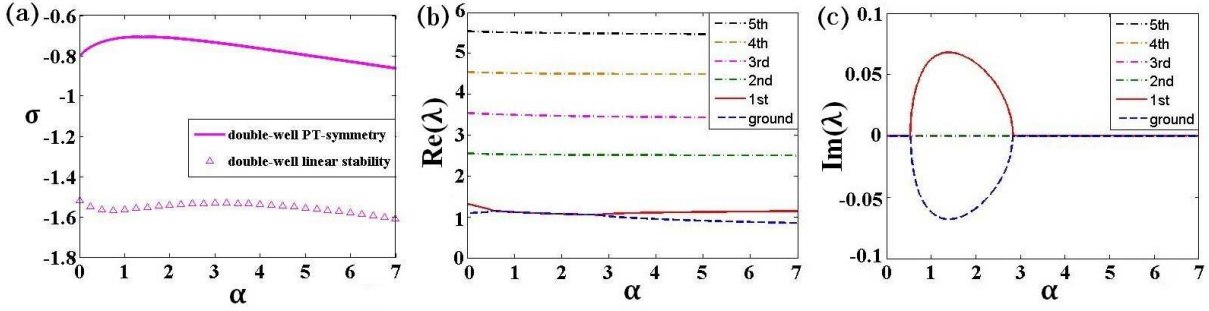}}}
 	\end{center}
 	\vspace{-0.15in} \caption{ (Color online). (a)  The line of the $\PT$-symmetry phase transition for the  $\PT$-symmetric potentials (\ref{dw}). The unbroken (broken) $\PT$-symmetric phase is in the domain above (below) the phase breaking lines. Triangle-shaped curve indicates the border of linear stability domain of the mode  (\ref{solu-dw}) with $n=0$. Stable (unstable) solitons are in the domains above (below) the respective border lines.  Real (b) and imaginary (c) parts of the eigenvalue $\lambda$ [see Eq.~(\ref{linearPT2})] as functions of $\alpha$ for the potential (\ref{dw}) at $\sigma=-0.73$.  In all panels $\epsilon=0.8$.}
 		\label{figdw}
 \end{figure*}

Unlike in the Case 1 considered in the previous Section, for the case of a multi-well potential the parameter $\sigma$ [see Eqs.~(\ref{poten3}) and (\ref{v2n})] need not to be non-positive and can acquire positive values, i.e. the nonlinearity can be either attractive or repulsive. Since however in this section we are interested in a double-well potentials, we concentrate on the ``deformation" of the linear potential
\bee\label{dw}
U_{\epsilon}(x)=\frac{x^2}{2}+\epsilon^2V_{10}(x)+\epsilon^4V_{20}(x)+i\epsilon W_0(x),
\ene
where
\bes\bee
\label{dp1} V_{10}(x)=-\frac{1}{2}e^{- x^2}, \quad
V_{20}(x)=-2\sigma e^{-(\alpha+1) x^2}, \qquad
\\
\label{dp3}  W_0(x)=-\frac{3}{2} x e^{-x^2/2}, \qquad
\ene\ees
[this corresponds to the chemical potential $\mu=\tilde{\mu}_0=1/2$] and require $\sigma<-1/(4\epsilon^2)$, the latter constraint ensuring the required double-well shape.

Notice that at $\alpha=0$ the nonlinearity is a constant [see Eq.~(\ref{poten3})] while the real part of the potential $U_{\eps}(x)$ preserves a  double-well shape. Nonlinear modes in this case were recently reported in~\cite{Cart,Midya} (also, nonlinear modes in a slightly different double well potential with the imaginary part (\ref{dp3}) was investigated in~\cite{Rodrig}). When $\alpha$ increases of the amplitude of the potential barrier between the humps  decreases.

As above we start by defining the domains of the unbroken and broken $\PT$-symmetric phase of the underline linear problem with $n=0$. Respectively, we consider the linear spectral problem [cf. Eq.~(\ref{linearPT})]
\begin{eqnarray}
\label{linearPT2}
\tilde{L}\Psi=\lambda\Psi,\,\, \tilde{L}=L_0\!+\!\eps^2 V_{10}(x)\!+\!\eps^4 V_{20}(x)\!+\!i\eps W_0(x),\,\,\,\,\,\,\,
\end{eqnarray}
where $L_0$ is given by Eq.~(\ref{lin}), $\lambda$ and $\Psi=\Psi(x)$ are the eigenvalue and eigenfunction, respectively. 
For the present case it will be important that the nonlinearity can also be varied in time and its value affects the stability of the mode. Indeed, in Fig.~\ref{figdw} (a) we show the domains of broken and unbroken phase at $\eps=0.8$ on the $(\alpha,\sigma)$-plane.  Parametric dependence of the lowest eigenvalues is shown in Figs.~\ref{figdw} (b) and (c). Like in the case of one-hump potential we observe that the unbroken phase corresponds to relatively large $\alpha$ and the spontaneous symmetry breaking occurs as $\alpha$ decreases (what corresponds to the increase of the width of the gain-and-loss domains) due to the collision of the two lowest eigenvalues. However now the broken phase corresponds to a limited interval of $\alpha$ and we observe the re-entered unbroken phase as $\alpha$ approaches zero.  We also notice that the instability is oscillatory:  the emergent complex eigenvalues have nonzero real part.

\begin{figure*}[!ht]
	\begin{center}
		\vspace{0.1in}{\scalebox{0.65}[0.65]{\includegraphics{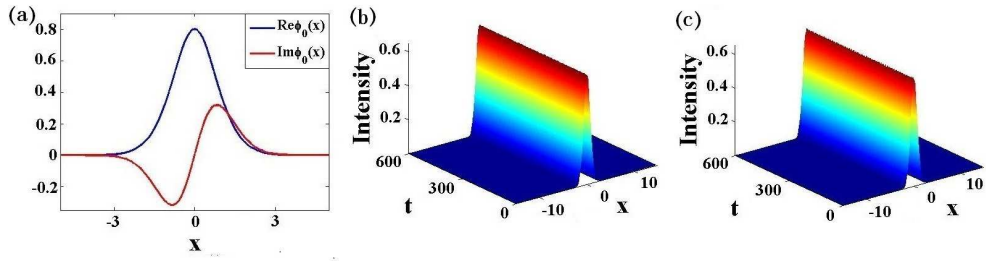}}}
	\end{center}
	\vspace{-0.2in} \caption{\small (Color online). The same nonlinear modes given by Eq.~(\ref{solu-dw}) with $n=0$,  $\alpha=4,\, \epsilon=0.8$ for both $\sigma=-0.65$ (real spectrum of the operator $\tilde{L}$, i.e. unbroken linear $\PT$-symmetry)	and  $\sigma=-1.25$ (broken linear $\PT$-symmetry) (a).
 Both nonlinear modes are linearly stable [see Fig.~\ref{figdw}(a)] what are illustrated in panels (b) and (c) where the direct numerical simulations of Eq.~(\ref{solu-dw}) are performed with weakly perturbed initial profile (a) for different parameters $\sigma=-0.65$ and $\sigma=-1.25$, respectively.} \label{fig9}
\end{figure*}

\subsection{Nonlinear modes in a double-well potential}

Now we turn to the nonlinear modes in the $\PT$-symmetric double-well potential (\ref{dw}), whose explicit expression is obtained from Eqs.~(\ref{ansatz_0}) and (\ref{vn2d}):
\bee \label{solu-dw}
\phi_n(x)=\eps H_n(x)e^{-x^2/2}\exp\!\left(\!i\eps\!\int_{-\infty}^x\!\!H_n(\xi)e^{-\xi^2/2}d\xi\!\right). \quad
\ene

The results of the linear stability analysis of solution (\ref{solu-dw}) are shown in Fig.~\ref{figdw}(a). The feature most relevant for our consideration consist in the domain, now with respect to $\sigma$, where the linear $\PT$-symmetry is broken, while the nonlinear mode remains stable (in analogy with the case of the one-well potential).

The stability of the nonlinear mode (\ref{solu-dw}) with $n=0$ is also confirmed by the direct propagation with the perturbation of initial profile, as illustrated in Fig.~\ref{fig9}. In Fig.~\ref{fig9} (b) we show stable propagation of the soliton for the parameters belonging to the domain of the unbroken linear $\PT$-symmetric phase  of the operator $\tilde{L}$ [defined in Eq.~(\ref{linearPT2})], and to the linearly stable nonlinear mode. In Fig.~\ref{fig9} (c) we illustrate the evolution of the mode where the linear $\PT$-symmetric phase is broken, however the nonlinear mode is linearly stable.

Now we turn to the excitation of nonlinear modes in the double-well potential. Since the potential $V_{20}(x)$ and nonlinearity $G(x)$ both contain $\sigma$, this parameter $\sigma$ can be exploited for managing (i.e. for excitation, in our case) of the nonlinear modes. To this end we  consider $\sigma$ to be a function of $t$ and address the
adiabatic switch-on of the potential and the nonlinearity, governed by the model (\ref{nls}) which now is rewritten in the form
\bee\label{nls-d2}
i\partial_t\psi=-\frac12\partial_x^2\psi+[V_0 (x)+\epsilon^2 V_{10}(x)+\epsilon^4V_{20}(x;\sigma(t)) \nonumber \quad\\
+i\epsilon W_0(x)]\psi+\epsilon^2G(x;\sigma(t))|\psi|^2\psi.\qquad\qquad
\ene
Here the double-well potential $V_{10}(x)$ and gain-and-loss distribution $W_0(x)$ are given by Eqs.~(\ref{dp1}) and (\ref{dp3}), respectively,  and $V_{20}(x;\sigma(t))$ $G(x;\sigma(t))$ stand for $V_{20}(x)$ and $G(x)$ given by Eqs.~(\ref{dp1}) and (\ref{poten3}) with $\sigma$ replaced by $\sigma(t)$.

\begin{figure}[!tp]
 	\begin{center}
 		\vspace{0.1in}
 \hspace{0.15in}{\scalebox{0.42}[0.5]{\includegraphics{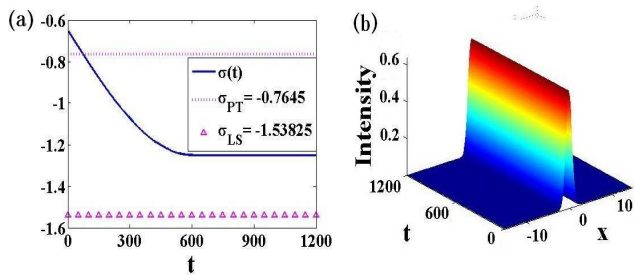}}}
 	\end{center}
 	\vspace{-0.15in} \caption{ (Color online).   (a) Time dependence of the parameter $\sigma(t)$ given by Eq.~(\ref{dwell}) (solid line). Upper dashed line and triangles indicate boundaries the stability of the nonlinear mode and the unbroken $\PT$-symmetric phase (in both cases above the respective curve). (b) Intensity evolution of nonlinear mode of Eq.~(\ref{nls-d2}) with the initial condition given by Eq.~(\ref{solu-dw}) for the double-well potential with $n=0$, $\alpha=4$,  and $\epsilon=0.8$. } \label{fig10-dwell-t}
 \end{figure}

According the general idea described above, now we choose the {\it adiabatic change} of $\sigma(t)$ in such a way that it assures that the system evolves from the domain of the unbroken $\PT$-symmetry of the underlying linear model to a broken phase, where however the nonlinear mode is linearly stable. More specifically, in the numerical simulations we exploit the dependence
\begin{eqnarray}\label{dwell}
\sigma(t)=\left\{
\begin{array}{ll} -0.6\sin\left(\dfrac{\pi t}{1200}\right)-0.65, & 0\leq t<600, \vspace{0.1in}\\
-1.25, & 600\leq t\leq 1200.
\end{array}
\right.\quad
\end{eqnarray}
The dependence $\sigma(t)$ is illustrated in Fig.~\ref{fig10-dwell-t} (a).

Fig.~\ref{fig10-dwell-t} (b) exhibits the excitation of a nonlinear mode described by Eq.~(\ref{nls-d2}) for the double-well potential, i.e. $V_0(x)$, $V_{10}(x)$, $V_{20}(x;\sigma(t))$, nonlinearity $G(x;\sigma(t))$ and gain-and-loss distribution $W_0(x)$ determined by Eq.~(\ref{dwell}) for the varying parameter $\sigma(t)$. The initial condition in these simulations is taken the form  (\ref{solu-dw}) with $n=0$. In the figure we again observe the stable evolution of the nonlinear mode between the initial and final shapes of the potential (notice that the change of the mode amplitude is relatively small and not clearly visible on the scale of the figure).\\

\section{Discussion and Conclusions}

In the present paper we reported several new branches of nonlinear modes described by the nonlinear Schr\"odinger equation with the $\mathcal{PT}$-symmetric single-well and double-well potentials and Hermite-Gaussian distributions of the gains and losses. The reported solutions are two-parametric, each of the parameters defining a branch of the solutions. All the considered branches bifurcate from the modes of the respective linear potentials. A peculiarity of the reported modes consists in their stability properties: their stability in the parameter space extends beyond the domains of the stability of the respective limits, i.e. beyond the domain of the unbroken $\PT$-symmetric phase of the underlying linear problem. This suggests a possibility of how a stable nonlinear mode can be excited in a $\PT$-symmetric system with broken $\PT$-symmetry. The method is based on the drawing the mode by adiabatic change of one of the control parameters along the branch from the domain where linear stability is verified and the $\PT$-symmetric phase is unbroken, i.e. where the mode can be excited by one of the conventional methods, to the domain where the $\PT$-symmetry is broken but the nonlinear stability still persists.

While the described approach is straightforward mathematically it leaves an open problem on managing gains and losses by a single parameter, while in many cases the physical phenomena responsible for  gains and losses have different nature. Speaking more generally, to implement the suggested scheme in practice one needn't to hold an integrable model but one does need to preserve balance between the gains and losses. This last goal be achieved in at least one of the ways as follows. First, considering $\PT$-symmetric profiles cerated in mixtures of atomic gases~\cite{gases}, one can modify both active and lossy domains by a single parameter, which is the intensity or wavelength of the control field, or alternatively by varying in space the properties of cladding of the atomic cell. The latter affects the whole spectrum of the underlying linear problem, i.e. both real and imaginary parts of the refractive index. On the other hand when considering $\PT$-symmetric double-well potentials in the BEC problems, to ensure the balance between varying gains and losses one has to perform simultaneous loading atoms to one of the wells (either using atomic laser~\cite{atom_laser} or tunneling from a neighbor potential well, as suggested in~\cite{Four-well}) and eliminating atoms (using, for example, ionization by external beam~\cite{Ott} or leakage of atoms through tunneling~\cite{Four-well}).

Finally, returning to the considered exact model, the method was tested using the found exact solutions in the form of one-- and two--hump modes supported by the one-well $\PT$-symmetric potential and by the repulsive nonlinearity, as well as for the modes in a double-well potential and attractive nonlinearity. In all the cases we observed the stable evolution of nonlinear modes, thus supporting practical feasibility of the approach.

\acknowledgments

Z.Y. and Z.W. were partially supported by the NSFC (No.61178091) and NKBRPC (No.2011CB302400). V.V.K. acknowledges support of the FCT (Portugal) grants UID/FIS/00618/2013 and PTDC/FIS-OPT/1918/2012.




\begin{references}

\bibitem{Bender98} C. M. Bender and S. Boettcher, Phys. Rev. Lett. {\bf 80}, 5243 (1998).

\bibitem{Bender2} C. M. Bender,
Rep. Prog.  Phys. {\bf 70}, 947 (2007).

\bibitem{Kato} T. Kato, {\em Perturbation Theory for Linear Operators} (Springer-Verlag, Berlin, 1980); M. Znojil, Phys. Lett, A {\bf 259}, 220 (1999).


\bibitem{sh} Z. Ahmed, Phys. Lett. A {\bf 282}, 343 (2001); {\em ibidem} {\bf 287}, 295 (2001).

\bibitem{Levai} G. L\'evai and E. Magyari, J. Phys. A: Math. Theor. {\bf 42}, 195302 (2009).

\bibitem{ZK} D. A. Zezyulin and V. V. Konotop,
Phys. Rev. A  {\bf 85}, 043840 (2012).

\bibitem{Musslimani} Z. H. Musslimani, K. G. Makris, R. El-Ganainy, and D. N. Christodoulides, Phys. Rev. Lett. {\bf 100}, 030402 (2008).

\bibitem{SII} H. Chen,  S. Hu,   and L. Qi, Opt. Commun. {\bf 331}, 139 (2014);  Z. Shi, X.  Jiang, X. Zhu,  and H.  Li,
Phys. Rev. A  \textbf{84}, 053855 (2011).



\bibitem{RMII} B. Midya and R.
Roychoudhury,
Phys. Rev. A {\bf 87}, 045803 (2013).


\bibitem{ZK_PRL} D. A. Zezyulin and V. V. Konotop,
Phys. Rev. Lett. {\bf 108}, 213906 (2012).


\bibitem{Segev} Y. Lumer, Y. Plotnik, M. Rechtsman, and M. Segev,
Phys. Rev. Lett. {\bf 111}, 263901 (2013).


\bibitem{AKSY} F. K. Abdullaev, V. V. Konotop, M. Salerno, and A. V. Yulin,
Phys. Rev. E {\bf 82}, 056606 (2010).

\bibitem{HP} Z. Y. Yan, {\it et al.,} arXiv:1009.4023;  Z. Y. Yan,  Phil. Trans. R. Soc. A {\bf 371}, 20120059 (2013).


\bibitem{gases} C. Hang,  G. Huang, and V. V. Konotop,
Phys. Rev. Lett. {\bf 110}, 083604 (2013);
C. Hang, D. A. Zezyulin, V. V. Konotop, and G. Huang,
Opt. Lett. {\bf 38}, 4033 (2013);
H. Li, J. Dou, and G. Huang,
Opt. Express, {\bf 21}, 108 (2013);
C. Hang, D. A. Zezyulin, G. Huang, V. V. Konotop, and B. A. Malomed,
Opt. Lett. {\bf 39}, 5387 (2014).

\bibitem{BECs} S. Klaiman, U. G\"unther, and N. Moiseyev,
Phys. Rev. Lett. {\bf 101}, 080402 (2008);
H. Cartarius and G. Wunner,
Phys. Rev. A, {\bf 86}, 013612 (2012).

\bibitem{Four-well}
M. Kreibich, J. Main, H. Cartarius, and G. Wunner,
Phys. Rev. A {\bf 87}, 051601(2013).

\bibitem{Cart}  D. Dast, D. Haag, H. Cartarius, J. Main, and G. Wunner, J.
Phys. A: Math. Theor. \textbf{46}, 375301 (2013).



\bibitem{Muga} A. Ruschhaupt, F. Delgado, and M. G. Muga,
J. Phys. A: Math. Gen. {\bf 38}, L171 (2005).


\bibitem{AKKS} F. K. Abdullaev,  Y. V. Kartashov, V. V. Konotop, and D. A. Zezyulin,
Phys. Rev. A {\bf 83}, 041805 (2011); D. A. Zezyulin,  Y. V.  Kartashov, and V. V. Konotop  Europhys. Lett.  {\bf 96}, 64003 (2011).

\bibitem{Michal} Y. He, X. Zhu, D. Mihalache, J. Liu, and Z. Chen, Phys. Rev. A {\bf 85}, 013831 (2012).


\bibitem{hm} J. Belmonte-Beitia, V.
M. Perez-Garcia, V. Vekslerchik, and P. J. Torres, Phys. Rev. Lett. {\bf 98}, 064102 (2007); V. N. Serkin, A. Hasegawa, and T. L. Belyaeva, Phys. Rev. Lett. {\bf 98}, 074102 (2007).

\bibitem{bb08} J. Belmonte-Beitia, V. M. P\'erez-Garc\'ia, V. Vekslerchik, and V. V. Konotop, Phys. Rev. Lett. {\bf 100}, 164102 (2008).

\bibitem{yan10} H. Friedrich, G. Jacoby, and C. G. Meister, Phys. Rev. A {\bf 65}, 032902 (2002);
Yu. V. Bludov,  Z. Y. Yan,  V. V. Konotop, Phys. Rev. A {\bf 81}, 063610 (2010).

\bibitem{yan09} Z. Y. Yan and V. V. Konotop, Phys. Rev. E {\bf 80}, 036607 (2009).

\bibitem{yan12} Z. Y. Yan and D. M. Jiang, Phys. Rev. E {\bf 85}, 056608 (2012); Z. Y. Yan, Stud. Appl. Math.
 {\bf 132}, 266 (2014).


\bibitem{bec} L. Pitaevskii, S. Stringari, {\em Bose-Einstein Condensation} (Oxford University Press, Oxford, 2003);
 C. J. Pethick, H. Smith, {\em Bose-Einstein Condensation in Dilute Gases} (2nd Edition). Cambridge University Press, Cambridge, 2008);
 Y. Kartashov, B. A.  Malomed, L. Torner, Rev. Mod. Phys. {\bf 83}, 247 (2011).


\bibitem{PTR} J. G. Muga, J. P. Palao, B. Navarro, and I. L. Egusquiza, Phys. Rep. {\bf 395}, 357 (2004).

\bibitem{CGS} E. Caliceti, S. Graffi and J. Sj\"ostrand  J. Phys. A: Math. Gen. {\bf 38},   185 (2005).

\bibitem{Rodrig}  A. S. Rodrigues, K. Li, V. Achilleos, P. G. Kevrekidis, D. J.
Frantzeskakis, and C. M. Bender, Rom. Rep. Phys. \textbf{65},
5 (2013).

\bibitem{Midya} B. Midya, Nonlin. Dyn. (2014) DOI 10.1007/s11071-014-1674-9

\bibitem{ss} J. A. C. Weideman, J. Phys. A {\bf 39}, 10229 (2006).

\bibitem{atom_laser}     N. P. Robins, P. A. Altin, J. E. Debs, and J. D. Close,
Phys. Rep. {\bf 529},  265 (2013)

\bibitem{Ott} G. Barontini, R. Labouvie, F. Stubenrauch, A. Vogler, V. Guarrera, and H. Ott,
Phys. Rev. Lett. {\bf 110}, 035302 (2013)


\end{references}
\end{document}